\documentclass{ab}  
\usepackage{graphicx,ulem}
\usepackage{txfonts}
\usepackage{color}
\begin{document}
  \titlerunning{ SS 433 O I emission}
\authorrunning{ Bowler}
   \title{SS 433: O I lines and the circumbinary disk }

   \subtitle{}

   \author{M. G.\ Bowler \inst{}}

   \offprints{\\   \email{michael.bowler@physics.ox.ac.uk}}
   \institute{University of Oxford, Department of Physics, Keble Road,
              Oxford, OX1 3RH, UK}

 
  \abstract
   {The Galactic microquasar SS\,433 may be the only local ULX.   The stationary Balmer H$\alpha$ and He I lines have two horned structures, interpreted as emission from a circumbinary disk orbiting a system of total mass approximately 40 $M_\odot$. The implication is that the compact object is a rather massive black hole.} 
  {To place on record the results of observations of two O I emission lines in the spectra of SS\,433; observations that support not only the presence of a circumbinary disk orbiting the system but also indicate its stability during flare episodes.  }
   {Stationary H$\alpha$ and He I spectra were analysed some years ago and are in excellent agreement with radiation from a circumbinary disk. All the evidence so far presented was adduced during periods in which SS\, 433 was quiescent; there has been little concerning the existence or state of the circumbinary disk during the violent episodes associated with radio flares. The behaviour of O I emission lines before and during such an outburst is addressed.  }  
  {The 8446 \AA\ O I line has a two horned structure and the two components of the split run railroad straight over some eighty days, in the latter half of which there were violent changes in H$\alpha$ and He I emission, culminating in a radio flare.}
 {O I 7772 \AA\ is mostly formed in a dense equatorial wind but  O I 8446 \AA\ is fluoresced in the circumbinary disk. That disk is not 
perturbed sensibly by the violent behaviour in the vicinity of the compact object. O I 8446 \AA\ acts as a tracer for the stable circumbinary disk orbiting at over 200 km s$^{-1}$, far outside the binary system.}

   \keywords{stars: individual: SS 433 - binaries: close - stars: fundamental parameters - circumstellar matter}

   \maketitle
%

\section{Introduction}

The  microquasar  SS 433 is unique in the Galaxy with continual ejection
of plasma in two opposite jets at approximately one quarter the speed
of light, though evidence for another such system ejecting baryonic plasma, in  M 81, has recently been presented (Liu et al 2015). The SS 433 system is a 13 day binary, powered by supercritcal accretion on the compact member. There is resurgent interest in SS 433 because it may be the only Ultra Luminous X-ray source in the Galaxy (Fabrika et al 2015; Kabibullin \& Sazonov 2015, Sadowski \& Narayan 2015, King \& Muldrew 2016). An important question is whether the compact object is a black hole and if so, a low or high mass stellar black hole. 

   Absorption spectra taken during and after primary eclipse, with the accretion disk turned most towards us, have been interpreted as formed in the atmosphere of an A type companion orbiting with a speed of approximately 60 km s$^{-1}$. The compact object would then be a neutron star or low mass black hole (Hillwig \& Gies 2008, Kubota et al 2010). Conflicting evidence is found in analysis of the stationary emission lines of hydrogen and helium. These have  two horned structures characteristic of emission from a disk seen close to edge on and do not share the orbital motion of the compact object. If the two horned features are emitted from a circumbinary disk, then that disk is orbiting the SS 433 system at over 200 km s$^{-1}$ and the mass of the compact object exceeds 16 $M_\odot$ (Blundell, Bowler \& Schmidtobreick 2008). A model in which intense radiation from the environs of the compact object ionises material orbiting in the circumbinary ring accounts for remarkably detailed features of the H$\alpha$ and He I spectra shown in Fig.2 of Schmidtobreick \& Blundell (2006b) and there may be other possible sources for the absorption spectra in the blue. A thorough discussion of the evidence for a circumbinary disk and its implications for the mass of the compact object can be found in Bowler (2013) and references given therein. Nonetheless, the more the two horned structure in emission lines is understood the better and there is one deficiency in the evidence. The two horned structures indicating a circumbinary disk orbiting at over 200 km s$^{-1}$ have only been studied during periods when SS 433 was quiescent and even if present during flaring episodes could not be extracted reliably. Are they present and unchanged but merely concealed or is the circumbinary source (if so it be) disrupted and later reformed? Here I discuss evidence from O I emission lines that there is indeed a circumbinary disk and that it survives unscathed a violent episode following an earlier period of quiescence.

\section{The relevant observations}

All the spectra discussed in this note are drawn from the remarkable data set of Schmidtobreick and Blundell. The observations with the ESO 3.8 m telescope on La Silla covered the period from JD 2453000 +245 to +321. Between + 245 and +274 spectra were taken nightly and only +252 was missed. Up to +274 SS 433 was quiescent and the stationary lines could be fitted in most cases with a superposition of three Gaussians, assigned to emission from a fast wind above the accretion disk and to a circumbinary disk. Observations were less continuous after +274, spectra being taken on +281, 282, 287 to 303 and 305 to 310. There was also an outlier at +321, after which SS 433 became a daylight object. After +274 the optical spectra broadened considerably and developed more complex structure. These optical phenomena culminated in a radio flare (Blundell et al 2011).

\subsection{ The circumbinary disk in H$\alpha$}

The two horned structure attributed to a circumbinary disk was originally detected in H$\alpha$, where the two horns maintained constant separation between days +245 and +274, running railroad straight (Blundell, Bowler \& Schmidtobreick 2008). The He I lines are a little more complicated. In H$\alpha$ the tracks become less regular between +281 and +287 and then on +288 a major morphological change occurs. A pair of Gaussian components separated by approximately 1000 km s$^{-1}$ appears and persists through to +321. The inner lines that might be continuations of the circumbinary disk become erratic, move to the red and on occasion disappear. This is illustrated in Fig.2 of Blundell et al ( 2011); I found much the same features in my independent fits to data after +287 (Bowler 2010). The newly emergent pair of lines share much of the orbital motion of the compact object (Blundell et al 2011, Bowler 2010), as does a broad Gaussian component originating in the wind above the accretion disk (Blundell et al 2011, Bowler 2011b). The question about the state of the observations of the circumbinary disk signature after +274 was summarised as follows in Blundell et al (2011), section 3.2:  "The disruption of the circumbinary disc: In the days that follow the flare, after accounting for a broad component of the H$\alpha$ complex as wind and the high velocity components as the manifestation of the accretion disc, in most cases there are one or two lines unaccounted for. It seems likely that some of these are the continuation of the signature of the circumbinary disc (albeit at different wavelengths from what is observed prior to Day +294), having possibly undergone some disruption during the the flare outburst. We assume that ...... this structure will regroup and form a more cogent signature as it has presumably  done following major flares in the past".

\subsection{The circumbinary disk in other emission lines}

In Fig.2 of Schmidtobreick \& Blundell (2006b) are exhibited the spectra day by day of H$\alpha$, He I 6678 \AA\,, He I 7065 \AA\ and the region of the oxygen line O I at 7772 \AA\ . The H and He I spectra beyond +287 are very broad and complex and exhibit features almost certainly absorption troughs. (Absorption of H$\alpha$ in a slow equatorial wind might account for the erratic shifts to the red in the supposed signature of the circumbinary disk.) O I 7772 \AA\ spectra are very different, presenting a clean P Cygni profile with particularly deep absorption troughs to the blue when the accretion disk is close to edge on. These troughs shift in wavelength with the orbital motion of the compact object and so the line O I 7772 \AA\ is likely to be formed (and certainly absorbed) in the slow equatorial wind (Bowler 2011b). Those spectra do not suggest any components from the high speed wind above the disk or from the accretion disk itself.

  Thus there is little hope of gaining information about the state of the circumbinary disk during violent episodes from any of these lines. However, there is a second O I line at 8446 \AA\ and this is split in the same way as the H$\alpha$ line is before +274. The daily spectra have not been published but I have access to the relevant files. This emission line has a simple split profile and the two components run railroad straight (like H$\alpha$ up to +274) from +245 to +321, unperturbed through the disturbances associated with the flare.
  
  \subsection{The circumbinary disk in O I 8446 \AA\ }
  
  In Fig.1 of Schmidtobreick \& Blundell (2006a) the full stationary spectrum of SS 433, after removal of the moving lines, is displayed. This is not the spectrum for a single day but rather the average over all daily spectra. It is nonetheless vivid and informative. In this average spectrum, O I 7772 \AA\ appears as a feeble line with a marked P Cygni profile. The height of the emission side is about 1/8 of the heights of the twin emission peaks at 8446 \AA\ . There is no indication of absorption to the blue in O I 8446 \AA\,, even though it is embedded in the Paschen series that does show absorption in this averaged spectrum. Fig.2 of Schmidtobreick \& Blundell (2006a) shows narrow regions for a single day, +274. The first two panels show the profiles of the H${\alpha}$  and He I 6678 \AA\ lines, showing the familiar twin peaks in each case. The third panel shows O I 8446 \AA\ and it has a profile very similar to the previous two. There is some minor interference from Paschen lines 14, 15 and 16; the Paschen 11 profile is shown in the fourth panel. [The convention is that employed in Schmidtobreick \& Blundell 2006a; Paschen 11 is the transition from principal quantum number 14 and may be labelled pa-$\lambda$.] The O I line is between Paschen 14 (to the red) and 16 (to the blue). Paschen 15 contributes a bulge on the blue side of the split O I 8446 \AA\ profile. The Paschen lines are a nuisance but fortunately the O I line is strong and the principal features clear. Day +274 is very close to the eclipse of the companion by the compact object and its accretion disc and it is clear from Fig.2 of Schmidtobreick \& Blundell (2006a) that the profiles of H$\alpha$, He I and O I 8446 \AA\ are on that day very similar indeed. On closer inspection the He I structure is narrower than H$\alpha$ (a factor of about 0.85) and in terms of km s$^{-1}$ O I is about the same width as He I. It is the case that H$\alpha$ is unusually broad in the few days up to and including +274 
  (Blundell, Bowler \& Schmidtobreick  2008) but in general the velocity dispersion in He I is a little less than that in H$\alpha$ (Bowler 2011a).
  
  The daily spectra cover both O I lines over the whole period +245 to +321. The line at 7772 \AA\ , if present at all, cannot be disentangled from the red shifted moving lines until + 270 and is feeble through +274. It strengthens after +281, which is also where the H$\alpha$ and He I lines start broadening, and becomes stronger and broader as flare conditions develop, with very strong P Cygni absorption profiles.
  The O I 8446 \AA\ line is strong throughout the whole sequence. Between +245 and +274 the profiles are very similar to the H$\alpha$ profiles on the same day, but a little narrower. Where two peaks are reasonably well defined the separation is approximately 250 km s$^{-1}$. There is no periodic motion from red to blue and back (as there is for He I lines and the underlying Paschen lines); the lines are as fixed as H$\alpha$ over this period. O I 8446\AA\ must have its origin in the circumbinary disk. The important feature of the profiles is manifest from day +281 onwards. The H$\alpha$ and He I lines broaden and become much more complex as wind speed increases, the accretion disk becomes visible and P Cygni absorption develops. O I 8446\AA\  shows no signs of these phenomena. The two horned structure associated with the circumbinary disk becomes if anything more sharply defined than on +274 and the two peaks run as straight as a railroad from +281 to the last spectrum at +321. (The underlying Paschen lines do shift periodically from red to blue and as time goes on broaden and develop P Cygni profiles.) The separation of the two peaks is approximately constant at over 250 km s$^{-1}$. In short, the O I 8446 \AA\ spectra have all the characteristics of azimuthally rather symmetric emission from a circumbinary ring, orbiting the SS 433 system at over 200 km s$^{-1}$.
  
  \section{Discussion}
  
  The spectra described above make it very clear that the O I lines at 7772 \AA\ and 8446 \AA\ originate in very different places. The 7772 \AA\ line is comparatively feeble but stronger during the eruption of the H$\alpha$ and He I spectra. Like those lines, it exhibits P Cygni absorption but even more strongly. There are indications that it shares at least some part of the orbital motion of the compact object. The absorption is particularly strong when the accretion disk is close to edge on. It seems very likely that this line is produced in the slow, dense equatorial wind. The line at 8446 \AA\ is strong, persistent and mimics the H$\alpha$ profile up to day +274. Thereafter it shows no obvious sign of eruption and the twin horned structure appears day after day with the twin peaks unchanged in wavelength. The obvious conclusion is that O I 8446 \AA\ is formed in the circumbinary disk and that the circumbinary disk is neither disrupted nor significantly disturbed during the optical outburst accompanying the radio flaring. The difficulties in tracing the circumbinary disk lines during the outburst after +294 must be due to the increased H$\alpha$ complexity and absorption destroying the visibility of the circumbinary disk lines, either wholly or to the blue. This is a useful coda to section 3.2 of Blundell et al (2011). More than that, it implies that the source of O I 8446 \AA\ is well outside the violent immediate environs of the compact object and hence that the disk is indeed circumbinary. Analyses of the H$\alpha$ and He I spectra up to day +274 have already provided very strong evidence that SS 433 possesses a circumbinary disk orbiting the centre of mass of the system at over 200 km s$^{-1}$. The O I 8446 \AA\ profiles provide entirely independent evidence for this and demonstrate that this disk is robust against  interior outbursts. Cute.
  
    Two questions arise. The profiles produced in O I 8446 \AA\ indicate that emission of this spectral line is fairly uniform round the circumbinary ring. This is true for H$\alpha$ but not for He I, nor for the Paschen series. Why is it true for O I 8446 \AA\ ? The second question is how it comes about that 8446 \AA\ is much more intense that 7772 \AA\ and apparently formed in a completely different environment. There is probably a single answer. The O I line at 8446 \AA\ can be fluoresced by H Ly$\beta$. If H$\alpha$ is stimulated with close to azimuthal symmetry, it seems likely that Ly$\beta$ is stimulated round the ring with similar azimuthal symmetry. Given a sufficient density of atomic oxygen in its ground state the O I 8446 \AA\ profiles might be expected. A sufficiency of O I ground state evidently exists in the circumbinary ring, perhaps toward the outer edge.
    
    \section{Conclusions}
    
    The O I emission line at 8446 \AA\ has, between days +245 and +274, all the characteristics that in H$\alpha$ have been attributed to origin in a circumbinary disk. In contrast to H$\alpha$ and He I lines, the signature twin peak structure continues to exhibit the characteristics of emission approximately uniform in azimuth from a circumbinary disk, in all spectra beyond +274 to the final spectrum at +321. It is not affected by changes leading to optical outbursts in H and He I emission spectra, outbursts that share at least some of the orbital motion of the compact object. It is the extraction of the signature of the circumbinary disk in H$\alpha$ that is disrupted after day +280 and the circumbinary disk itself is little affected. This is additional evidence that the circumbinary disk is indeed circumbinary, with the usual corollary that the compact object is a rather massive black hole.

\begin{acknowledgements}
The spectra covering the O I emission lines day by day were made available to me over 7 years ago by K. M. Blundell. The observations were made in 2004 with the ESO NTT, a grant of Director's discretionary time.

\end{acknowledgements}

\end{document}